\documentclass[useAMS,usenatbib]{mn2e}
\usepackage{graphicx}
\title[Metallicity and Giant Planet Incidence]{The Metallicity Dependence of Giant Planet Incidence}
\author[G.\ Gonzalez]{G.\ Gonzalez$^{1}$\\
$^{1}$Department of Physics and Astronomy, Ball State University, Muncie, IN 47306 USA\\
}

\begin{document}

\date{Accepted ??. Received ??; in original form ??}

\pagerange{\pageref{firstpage}--\pageref{lastpage}} \pubyear{??}

\maketitle

\label{firstpage}

\begin{abstract}
We describe three corrections that should be applied to the observed relative incidence of nearby stars hosting giant planets. These are diffusion in the stellar atmosphere, use of the [Ref] index in place of [Fe/H] for metallicity, and correction for local sampling with the W velocity. We have applied these corrections to a subset of the SPOCS exoplanet survey with uniform giant planet detectability. Fitting the binned data to a power law of the form, $\alpha 10^{\beta [Fe/H]}$, we derived $\alpha = 0.022 \pm 0.007$ and $\beta = 3.0 \pm 0.5$; this value of $\beta$ is 50\% larger than the value determined by \citet{fv05}. While the statistical significance of this difference is marginal, given the small number statistics, these corrections should be included in future analyses that include larger samples.
\end{abstract}

\section{Introduction}

Since the correlation between the metallicity of a star and the presence of giant planets was first discovered \citep{gg97}, several research groups have worked to determine the precise functional form of the relationship \citep{fv05,mor13,nev13}. This relationship is important to know accurately, since it constrains planet formation models and simulations \citep{benz14} and allows us to predict giant planet incidence in regions of the universe with different metallicity distribuitons relative to the solar neighborhood.

In most stellar chemical abundance studies the measured metallicity of a star's atmosphere is equated with the star's initial bulk metallicity. It has been known for many years that this assumption is not strictly correct, due, primarily, to the effects of atomic diffusion in a star's atmosphere. However, the magnitude of the error in metallicity made from ignoring diffusion is small or insignificant for most stars. For this reason, it has been considered justifiable to ignore the effects of diffusion given that the measurement errors are usually larger.

Some problems in modern astrophysics require very precise stellar metallicity data. Two examples are the local G dwarf metallicity distribution for constraining Galactic chemical evolution and the metallicity dependence of giant planet incidence. Both these examples would benefit from corrections for diffusion. We will focus only on the second example in the present study. Previous work on this topic has revealed that the incidence of giant planets is a very sensitive function of the [Fe/H] value of the host star \citep{fv05}. However, the precise functional form of this relation remains uncertain \citep{mor13,nev13}. None of the prior studies corrected for the effects of diffusion.

There are two additional factors that should be considered in deriving the dependence of giant planet incidence on the metallicity of the host star. One is the definition of metallicity. \citet{gg09} introduced a new metallicity index, called ``[Ref]'', which is better suited to relating the compositon of the host star to the formation of giant planets. Another factor is the correction for the bias in the selection of the samples used to examine the metallicity dependence of giant planet incidence. The volume-limited surveys of nearby stars necessarily undersample stars that spend most of their time far from the Galactic mid-plane. This can be corrected easily with the Galactic W velocity value for each star \citep{wie96}.

The purpose of the present study is to revisit the problem of the metallicity dependence of giant planet incidence and attempt to improve on previous determinations. In Section 2 we describe the three corrections to the observed incidence of nearby giant planets. In Section 3 we apply the corrections to a subset of the SPOCS sample \citep{vf05}. We present our conclusions in Section 4.

\section{Corrections to planet incidence}

\subsection{Correcting for Diffusion}

\citet{mow12} produced dense grids of stellar models that include the effects of atomic diffusion. For each record in their grids they list both the initial metallicity and the surface metallicity of a star. From these tabulations we form the quantity, $\Delta[Fe/H]_{D} = [Fe/H]_{s} - [Fe/H]_{i}$, which is a quantitative measure of the alteration of the surface metallicity of a star by diffusion relative to its initial metallicity. We have taken 67 cases from their tabulations spanning the following ranges in mass, age, and initial [Fe/H] value ($[Fe/H]_{i}$), respectively: $0.7$ M$_{\odot} \le m < 1.1$ M$_{\odot}$, $0.5$ Gyr $<$ age $< \sim11$ Gyr, and $-0.33 \le [Fe/H]_{i} \le 0.54$. This range in [Fe/H] was selected from the available values tabulated by \citet{mow12} that encompasses the values [Fe/H] of the planet host stars in our sample. \citet{mow12} did not include diffusion for stars with masses $\ge 1.1$ M$_{\odot}$; for stars in this mass range they set the surface [Fe/H] value equal to the initial [Fe/H] value. Over these ranges of stellar parameters, $\Delta[Fe/H]_{D}$ ranges form near zero to about -0.10 dex. We fit the following simple polynomial multivariate equation to the $\Delta[Fe/H]_{D}$ values using the data from these 67 cases: 
\begin{eqnarray*}
\Delta[Fe/H]_{D} = \left\{ \begin{array}{ll}
- \vert~a + b m + c m^2 +&\mbox{ if $0.7 < m < 1.1$}\\
d \log Age + e \log^2 Age~+ \\
f \log^3 Age + g \log^4 Age~+ \\
h~[Fe/H]_{o} + i~[Fe/H]^{2}_{o}~\vert \\
\\
0 &\mbox{if $m \ge 1.1$}\\
 &\mbox{or $Age < 0.5$ Gyr}\\
 &\mbox{or $\log g < 4.0$}
       \end{array} \right.
\end{eqnarray*}
where $m$ is mass in units of M$_{\odot}$, $\log Age$ is the log (base 10) of the age of the star in years, and $[Fe/H]_{o}$ is the observed [Fe/H] value of the star's atmosphere (the current surface [Fe/H] value). The values of the coefficients $a-i$ are listed in Table 1. 

The average difference of the fitted $\Delta[Fe/H]_{D}$ values from the \citet{mow12} values is $5 \times 10^{-4}$ dex (in the sense of their values minus our interpolations); the the standard deviation of the differences is $7 \times 10^{-3}$ dex. Our interpolation equation captures the basic dependences of $\Delta[Fe/H]_{D}$ on mass, age and $[Fe/H]_{o}$. In particular, the magnitude of $\Delta[Fe/H]_{D}$ increases steadily with age while a star remains on the main sequence. The interpolation equation begins to break down for the older stars. As a star begins to leave the main sequence, the magnitude of $\Delta[Fe/H]_{D}$ begins to decline, but this effect is not reproduced by our interpolation equation. For this reason, we have set $\Delta[Fe/H]_{D}$ to zero for $\log g < 4.0$. In addition, since the magnitude of $\Delta[Fe/H]_{D}$ is less than about 0.01 dex for yougn stars, we have set $\Delta[Fe/H]_{D}$ to zero in the interpolation equation for age $< 0.5$ Gyr.

\begin{table*}
\centering
\begin{minipage}{160mm}
\caption{Values of the constants in the interpolation equation for $\Delta[Fe/H]_{D}$.}
\label{xmm}
\begin{tabular}{ll}
\hline
Constant & value\\
\hline
$a$ & ~557.965 \\
$b$ & ~0.562481 \\
$c$ & -0.359897 \\
$d$ & -238.141 \\
$e$ & ~38.0506 \\
$f$ & -2.69821 \\
$g$ & ~0.0716336 \\
$h$ & ~0.0404004 \\
$i$ & -0.0682419 \\

\hline
\end{tabular}
\end{minipage}
\end{table*}

In practice, the value of $[Fe/H]_{i}$ is calculated by subtracting the interpolated value of $\Delta[Fe/H]_{D}$ from the observed value of [Fe/H]. For example, the value of $\Delta[Fe/H]_{D}$ calculated from the interpolation equation above for the Sun is -0.03 dex, which gives a solar $[Fe/H]_{i}$ value of 0.03 dex. For stars other than the Sun, the masses and ages are determined from stellar models that do not usually include the effects of diffusion in relating the observed [Fe/H] of a star to $[Fe/H]_{i}$. For this reason, the mass and age values entered into the interpolation equation for $\Delta[Fe/H]_{D}$ will be slightly in error. However, since $\Delta[Fe/H]_{D}$ is a slowly varying function of mass, age and the observed [Fe/H] and since $\Delta[Fe/H]_{D}$ is small for the ranges in these parameters we consider, the errors committed in estimating $[Fe/H]_{i}$ should be small.

\subsection{Refractory Metallicity}

\citet{gg09} introduced a new metallicity index, [Ref], in order to explain an apparent preference of giant planets for thick disk and transition disk stars with [Fe/H] $\le -0.20$ dex as compared to thin disk stars over the same interval in [Fe/H], which had been reported by \citet{hay08}; in addition, \citet{hay09} suggested that giant planet formation must depend on Galactic location in order to account for this difference. However, use of the [Ref] index in place of [Fe/H] eliminates the apparent preference of giant planets for thick disk stars relative to thin disk stars for [Fe/H] $\le -0.20$ and, hence, any special dependence on Galactic location. This was also confirmed by \citet{adib12}.

The motivation for introducing the [Ref] index is as follows. Iron is not the only abundant refractory element in the Solar System; magnesium and silicon have comparable number abundances and condensation temperatures to iron \citep{lod09}. If we are interested in examining the elements that go into building giant planets, then we need to consider the sum of the masses of these three elements in a star. The [Ref] index to quantifies the mass abundances of Mg, Si and Fe relative to the Sun:

\begin{eqnarray*}
{\rm [Ref]}=\log(24\times10^{(7.55+{\rm [Mg/H]})}+28\times10^{(7.53+{\rm [Si/H]})}+\\
56\times10^{(7.47+{\rm [Fe/H]})})-9.538
\end{eqnarray*}

The name of the index, [Ref], borrows the bracket notation from the standard number abundance notation, since the index is logarithmic and the abundances are relative to the Solar System values. ''Ref'' is short for ''refractory.''

\subsection{Correcting for W Velocity}

Samples of nearby stars will necessarily undersample stars that reach large distances from the Galactic mid-plane due to their space motions. Thick disk stars and older thin disk stars will thus be undersampled in surveys of nearby stars. \citet{wie96} corrected for this selection bias by weighting metallicity values by $|$ W+W$_{\odot} |$. In our analysis below we adopt Wielen's method with a value of 7 km s$^{-1}$ for W$_{\odot}$, which is near the average of recent determinations \citep{schon10,fran14}.

\section{Applying corrections to the SPOCS dataset}

\citet{fv05} produced a histogram distribution of the incidence of giant planets as a function of [Fe/H] using a subsample of the 1040-star SPOCS dataset \citep{vf05}. It consists of 850 FGK stars with uniform abundance analyses and planet detectability (listed in their Table 3). To be included in their planet hosts sample, a star must satisfy three criteria: planet orbital period shorter than 4 yrs, K value greater than 30 m s$^{-1}$, and at least 10 observations over four years. Of the 850 stars in their subsample, 47 satisfy these criteria.

\citet{fv05} show the resulting histogram distributions in their Figure 4 and 5; Figure 4 displays the historgram with 0.25 dex-wide bins, while Figure 5 uses 0.1 dex-wide bins. They calculated the error bars for each bin according to standard Poisson counting statistics. They fit the distribution in their Figure 5  to a simple power law of the form:
\begin{eqnarray*}
\mathcal{P}({\rm planet})=\alpha \times 10^{\beta [Fe/H]} = 
\alpha \left[\frac{N_{Fe}/N_{H}}{(N_{Fe}/N_{H})_{\odot}}\right]^{\beta}
\end{eqnarray*}
They obtained $\alpha = 0.03$ and $\beta = 2.0$ (no error bars were given). We will determine new estimates for the $\alpha$ and $\beta$ parameters in the following.

Starting with the same 850 star subsample (and the 47 host star subsubsample) of SPOCS employed by \citet{fv05}, we will calculate new distributions of giant planet incidence based on the three corrections discussed in the previous section. In order to calculate the diffusion correction using the interpolation equation from Section 2.1, we first need to compile mass and age estimates for the stars. 

\citet{tak07} determined masses and ages for the SPOCS sample using Bayesian statistics. For some of the stars, they were not able to derive ages. We have adopted their mass and age esitmates with the following modifications. First, we deleted those stars lacking age constraints. If only a minimum age is quoted, and it is greater than 8 Gyr but less than 10 Gyr, then we set the age to 11 Gyr; if the minimum age is between 10 and 12 Gyr, we set the age to 12 Gyr; for minimum ages greater than 12 Gyr we set the age to the minimum age; if only a minimum age is given and it is less than 8 Gyr, we deleted the star. If only a maximum age is quoted and it is less than 3 Gyr, we set the age to half the maximum age; if only a maximum age is given and it is greater than 3 Gyr, we deleted the star. We also excluded stars with mass estimates less than 0.9, for which the age estimates tend to be less reliable. Although this procedure might seem inprecise, it is adequate for the purpose of calculating the diffusion metallicity corrections.

In order to apply the W velocity correction, we made use of the large database of space velocities from \citet{bob06}; a few stars not in their study were supplemented with data from the Simbad database. Our final cross-referenced subsample from the 850 star subsample of SPOCS contains 581 stars; it consists of 39 planet hosting stars and 542 comparison stars. We show the distributions of the stars with planets (SWPs) and comparison stars in Figure 1. The lowest value of [Fe/H] among the SWPs is -0.22 dex. Only one SWP and one comparison star are in the [Fe/H] = 0.5 - 0.6 dex bin; for this reason, we exclude this bin from the subsequent analyses. It is instructive to plot the SWP incidence distribution for each correction we apply in order to compare the relative importance of each one.

The average values of the age and mass estimates of our planet hosts subsample are $6.7 \pm 3.3$ Gyr and $1.14 \pm 0.20$ M$_{\odot}$, respectively. The corresponding values for the comparison stars subsample are $5.6 \pm 3.0$ Gyr and $1.10 \pm 0.16$ M$_{\odot}$, respectively. The larger average age for the planet hosts relative to the comparison stars implies that the diffusion and W velocity corrections will be somewhat greater for the planet hosts.

\begin{figure}
  \includegraphics[width=3in]{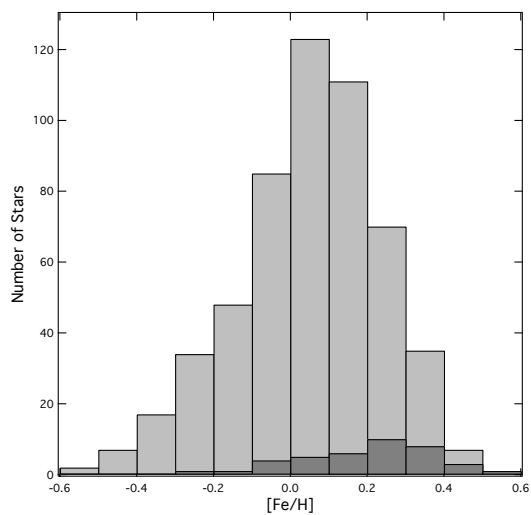}
 \caption{[Fe/H] distributions of the planet host and comparison stars.}
\end{figure}

In Figure 2 we display the fractional incidence of the planet host stars in each [Fe/H] bin from Figure 1. The error bars were calculated using the method described by \citet{cam11}; one-sigma upper and lower bounds are shown for each bin containing planet hosts. The distribution looks similar to the one from Figure 5 of \citet{fv05}. A weighted least-squares power-law fit to the data in Figure 2 yields $\alpha = 0.033 \pm 0.006$ and $\beta = 2.1 \pm 0.3$. The fit includes data from the [Fe/H] $= -0.3$ bin to the 0.5 bin, and the weight for each bin was set to the inverse square of the error, where the error here is the mean of the magnitudes of the lower and upper error bounds. If, instead, we calculate the error bars using propagation of errors and Poisson counting statistics (as Fischer and Valenti did in their study), we obtain $\alpha = 0.034 \pm 0.006$ and $\beta = 2.0 \pm 0.3$; these results are indistinguishable from those determined by them. This shows that we have not introduced significant biases in producing our subsample from their 850 star subsample of SPOCS.

In order to test the sensitivity of the results to the binning choices, we have repeated the above analysis (with the correct error bar calculations) with two changes to the binning. First, we changed the binning width from 0.1 to 0.15 dex; the best fit to this choice resulted in $\alpha = 0.031 \pm 0.009$ and $\beta = 2.1 \pm 0.5$. Second, we kept the bin width at 0.1 dex but shifted the bins by 0.05 dex; the best fit to this choice resulted in $\alpha = 0.029 \pm 0.006$ and $\beta = 2.3 \pm 0.3$. For both tests, the number of planet hosts included in analysis was the same as the nominal case. We conclude from these tests that the binning choices result in small changes to the solutions, which are smaller than the derived errors. 

\begin{figure}
  \includegraphics[width=3in]{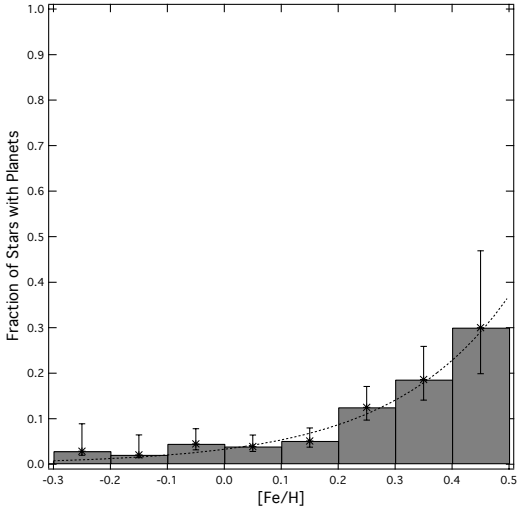}
 \caption{Relative incidence of SWPs with [Fe/H]. The dashed curve is the best-fit power law to the binned data; same in the subsequent figures.}
\end{figure}

In Figure 3 we show the SWP fractional incidence data corrected for diffusion; [Fe/H]$_{\rm D}$ is the observed [Fe/H] value corrected for diffusion using the interpolation equation from the previous section. The best-fit power-law to these data gives $\alpha = 0.023 \pm 0.006$ and $\beta = 2.6 \pm 0.4$. 

\begin{figure}
  \includegraphics[width=3in]{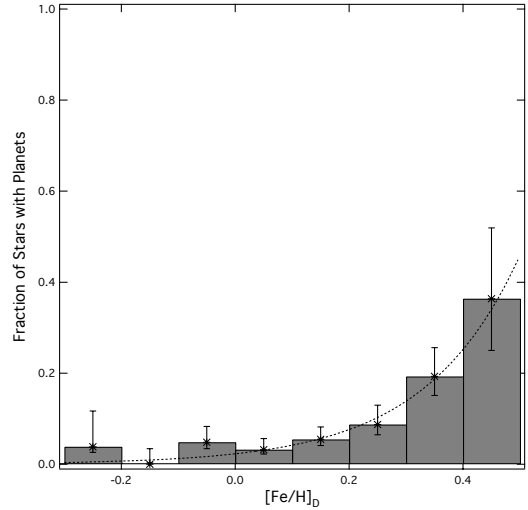}
 \caption{Relative incidence of SWPs with [Fe/H]$_{\rm D}$.}
\end{figure}

In Figure 4 we show the SWP fractional incidence distribution in terms [Ref]$_{\rm D}$, which is the [Ref] metallicity corrected for diffusion. Since the SPOCS stars lack [Mg/H] determinations, we have substituted [Ti/H] for [Mg/H] in the equaiton for [Ref] given in Section 2.2.\footnote{In principle, Si could be used as a substitute for Mg when calculating [Ref]. However, Ti is probably the better choice as a subsitute for Mg, since [Ti/Fe] behaves more like [Mg/Fe] among nearby stars than does [Si/Fe] when plotted against [Fe/H]. See Figure 8 of \citet{adib12} and Figure 15 of \citet{bens14} for a recent large nearby stars spectroscopic survey showing these trends.} In addition, we assume that the diffusion correction is the same for Si, Fe and Ti.  The best-fit power-law to these data gives $\alpha = 0.023 \pm 0.004$ and $\beta = 2.8 \pm 0.3$.

\begin{figure}
  \includegraphics[width=3in]{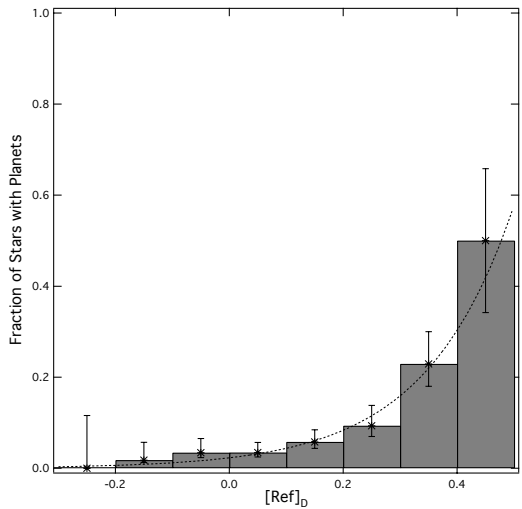}
 \caption{Relative incidence of SWPs with [Ref]$_{\rm D}$.}
\end{figure}

It is instructive to note that the large increase in the fractional incidence of stars with planets in the most metal-rich bin in going from Figure 4 to Figure 5 is due entirely to the decrease in the numbe of comparison stars in this bin (from 7 to 4 stars). This implies that the most metal-rich comparison stars tend to be more $\alpha$-element poor than the most metal-rich planet host stars. It would be worthwhile to test this finding with a larger sample of very metal-rich stars and different ways of calculating [Ref], e.g., using Mg instead of Ti.

Finally, in Figure 5 we show the SWP fractional incidence data in terms [Ref]$_{\rm DW}$, which is corrected for diffusion and the W velocity. The best-fit power-law to these data gives $\alpha = 0.022 \pm 0.007$ and $\beta = 3.0 \pm 0.5$. These results are consistent with the broad range of parameter values reported by \citet{nev13}, who did not apply these corrections to their data.

\begin{figure}
  \includegraphics[width=3in]{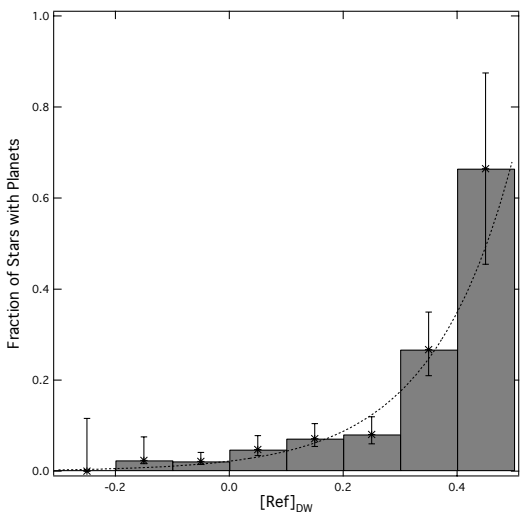}
 \caption{Incidence of SWPs with [Ref]$_{\rm DW}$.}
\end{figure}

\section{Conclusions}

We have applied three previously negelected corrections to the distribution of the local SWP fractional incidence. Application of each correction made a noticable difference to the distribution, though the W-velocity correction resulted in the smallest change. Our best fit to the distribution is modestly different from the one originally determined by \citet{fv05}; our fit shows a greater sensitivity to metallicity. In this process we have replaced the traditional index for metallicity, [Fe/H] or [M/H], with [Ref].

We advocate that these corrections be applied to any future determinations of the local SWP incidence. We also encourage the use of [Ref] in planet formation simulations in place of [Fe/H]. Use of [Ref] permits inclusion of thin disk, transition, and thick disk stars in the same sample. Finally, we encourage that both the measured [Fe/H] and the diffusion-corrected [Fe/H] values be reported for nearby stars included in detailed spectroscopic analyses. Our simple interpolation equation for the diffusion correction makes this an easy calculation.

\section*{Acknowledgments}

We thank the anonymous reviewer for helpful comments that improved the quality of this paper. This research has made use of the SIMBAD database, operated at CDS, Strasbourg, France.

\bsp

\label{lastpage}

\end{document}